# Energy-Efficient Ternary Encoding for High-Speed Data Transmission in 3D-Integrated Circuits Using Inductive Coupling Links

Abdullah Saeed Alghotmi
Computer Science Department
Faculty of Computing and Information
Al-Baha University, Saudi Arabia
aalghotmi@bu.edu.sa

*Abstract*— This paper proposes a ternary signalling scheme for inductive coupling links (ICLs) in 3D-integrated circuits (3D-ICs) to reduce crosstalk and electromagnetic interference in multi-stacked chip communications. By converting binary data into ternary sequences with three voltage levels (-V, 0V, +V), the approach enhances signal separation, reduces crosstalk, and improves signal integrity. Unlike traditional Non-Return to Zero (NRZ) systems, the ternary scheme increases bandwidth efficiency and reduces power consumption through fewer signal transitions. A modified H-Bridge transmitter generates ternary symbols by controlling current flow based on binary-to-ternary mapping. Preliminary simulations validate the efficiency of the scheme, showing reduced power consumption and higher data rates compared to NRZ. This approach shows promise for high-performance computing and IoT devices in 3D-IC environments, offering enhanced noise resilience, lower power usage, and improved communication efficiency.

## I. Introduction and Background

The continuous reduction in chip sizes and doubling of transistor capacity approximately every two years, as predicted by Moore's law, has driven research into multi-layered chip architecture [1]. Multi-chip (3D) systems are proposed to overcome the limitations of the conventional 2D architectures [2]. To vertically link stacked chips, there exist two approaches; wired-based and wireless-based approaches. The state-of-the-art of the wired-based approach is Through-silicon-via (TSV). However, this approach goes through several costly stages and prone to defects [ 2]. On the other hand, wireless-based approaches including capacitive and inductive coupling links have become more appealing due to their lower cost and high performance. The capacitive coupling links have some limitations due to technological integration while inductive coupling can communicate widely and not limited to coil design style. Inductive coupling links (ICLs) are particularly attractive for vertical communication due to their ease of alignment and cost-effectiveness [3]. ICLs have been applied in a variety of scenarios, such as SDRAM and IoT applications [ 4]. However, ICL suffers from the interference between neighbouring coils, which can disrupt timing and introduce noise. This interference manifests as jitter in the time domain and voltage noise in the frequency domain, or both [5]. This phenomenon is called crosstalk, caused by mutual inductance between adjacent coils. To mitigate crosstalk, several techniques have been developed, including increasing the physical distance between coils, which impacts manufacturing complexity [6], choosing encoding or a signalling schemes that aligns with the characteristics of inductive coupling [7], and employing time interleaving multiplexing [8].

3D-ICs using ICLs transfer data via magnetic fields generated by current pulses. When a pulse is sent from the transmitter (primary coil), the receiver (secondary coil) detects two complementary pulses, corresponding to the rise and fall of the primary. This process operates based on Faraday's Law of Induction, which explains how a changing magnetic field within a closed loop. Maxwell's equations provide a detailed framework for understanding the interactions of these electromagnetic fields.

Several signalling schemes are employed in ICL communication. For instance, bi-phase signalling represents binary data as a sequence of current pulses, assigning "1" to a positive pulse and "0" to negative pulse. Another common approach is Non-Return to Zero (NRZ), which is widely used in 3D-ICs. Unlike bi-phase, NRZ maps rising edges to positive pulses and falling edges to negative pulses. Both signalling schemes illustrated in Figure 1.

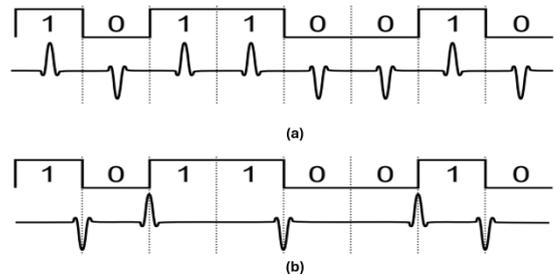

*Figure 1 Pulse representations in ICL communications using (a) Bi-phase, (b) Non-return-to-zero NRZ [9]*

As an alternative solution, this work explores a ternary signalling scheme with three voltage levels (-V, 0V, +V) designed to further minimise crosstalk by reducing current transitions and improving signal separation. Unlike traditional binary schemes, ternary encoding has the potential to enhance bandwidth efficiency and reduce power consumption, making it a promising approach for high-density 3D-IC communications. By leveraging this encoding scheme, we aim to address the limitations of conventional methods and offer a more robust solution for reducing interference in ICLs in 3D-ICs.

## II. RELATED WORK

The increasing density of 3D-ICs necessitates advanced vertical communication methods to ensure signal integrity, energy efficiency, and high data rates. ICLs offer alignment ease and cost advantages but face limitations with traditional signalling schemes like NRZ and bi-phase, particularly in high-density environments [2]. This section reviews key signalling techniques, the role of ICLs in 3D-ICs, and strategies for enhancing communication performance. Additionally, it examines the ternary signalling as a promising alternative to overcome these limitations by increasing data throughput and reducing interference.

### A. Crosstalk Mitigation Techniques

Recent advancements in mitigating crosstalk have concentrated on enhancing coil layout designs and utilising advanced signalling schemes. The authors in [9] investigated various signalling methods, including single-phase modulation (SPM), bi-phase modulation (BPM), and non-return to zero (NRZ) signalling, aimed at lowering power consumption in ICLs. Additionally, they introduced a phase code modulation technique that, while effectively reducing power usage, presents a lower resolution which may lead to timing uncertainties.

The impact of crosstalk on power distribution networks within inductive and capacitive coupling in 3D-ICs was examined in [3]. The authors employed both analytical and numerical approaches to model the mutual inductance between coils and suggested design modifications aimed at reducing crosstalk. Their research offered valuable insights into how coil separation and layout influence crosstalk levels.

Our previous work on crosstalk in inductive coupling communications for multi-stacked chips examined these challenges in high-density 3D-IC environments and proposed preliminary mitigation techniques to enhance data integrity [19]. This research offered an understanding of how electromagnetic interference could be minimized by adjusting coil layouts and managing interference across adjacent coils. Building on these findings, the current study explores a novel ternary signaling scheme to further reduce interference, minimizing transition rates to enhance both power efficiency and signal integrity in dense 3D-IC configurations.

### B. Energy Efficiency and Data Transmission in 3D-ICs

In multi-layered 3D-ICs using ICLs as vertical data transmission, energy efficiency and data transmission rates are essential for meeting the needs of high-performance computing and IoT applications. Traditional binary signalling methods, such as NRZ and bi-phase signalling, face limitations in complex 3D-IC using inductive coupling links. NRZ signalling consumes power due to continuous signal levels, which leads to higher energy usage in densely packed systems [9]. Similarly, bi-phase signalling, which requires more frequent transitions, further increases power consumption and limits data rates in high-density configurations [10].

To overcome these challenges, ternary signalling provides a promising alternative. By encoding data using three distinct voltage levels (-V, 0V, +V), ternary signaling reduces the frequency of transitions and thus lowers power consumption compared to traditional binary schemes. Furthermore, this multi-level signaling approach allows for higher data throughput and bandwidth efficiency in ICLs, addressing the constraints of conventional methods while maintaining signal integrity and reducing crosstalk in high-density environments.

### C. Signalling Techniques in 3D-ICs

Traditional signalling techniques, such as NRZ and bi-phase signalling, are widely used in 3D-ICs due to their simplicity and ease of implementation [11],[12]. NRZ signalling offers straightforward data representation with low complexity [13], while bi-phase signalling provides inherent clock recovery, which aids in synchronization [14].

However, as 3D-IC architecture becomes more complex and denser, these binary signalling schemes face significant limitations. NRZ, with its continuous signal levels, suffers from increased power consumption and crosstalk in multi-layered environments, while bi-phase requires more frequent transitions, which can limit its performance in high-density setups [15]. These challenges underscore the need for more advanced schemes that can provide higher data rates and improved energy efficiency without compromising signal integrity.

Prior studies have proposed ternary signaling schemes as a theoretical approach to enhance data throughput and power efficiency. For instance, the paper in [16] reviewed various ternary logic implementations, highlighting their potential in reducing power consumption and increasing data efficiency within high-density circuits, though they did not focus specifically on 3D-ICs or ICLs. Additionally, [17] outlined the potential advantages of employing three-level signaling in inductive coupling links (ICLs), suggesting that ternary encoding could improve data transmission rates and reduce power consumption compared to traditional binary methods. While these works primarily examined theoretical feasibility, practical implementation and validation were not pursued.

Building on this foundation, our study not only explores theoretical aspects of ternary signaling but also implements and validates its performance through simulations in Advanced Design System (ADS) and Ansys High-Frequency Structure Simulator (HFSS). By examining both on-chip and off-chip scenarios, we present empirical data on data transmission rate improvements and power efficiency gains, bridging the gap between theory and practical application in ternary ICL systems.

### D. Contribution

This paper proposes and presents a novel approach in 3D-ICs using ICLs to improve data transmission rates and power efficiency in 3D-ICs utilizing ICLs. The primary contributions of this work are as follows:

• **Enhanced data transmission through ternary signalling**: We introduced a ternary encoding scheme that uses three voltage levels to encode data, allowing for higher data throughput compared to traditional NRZ.

• **Improved power efficiency**: Reduced signal transitions in ternary signaling lead to lower power consumption in dense 3D-IC setups.

• **Crosstalk reduction**: Minimizes interference through reduced transition rates, enhancing signal integrity.

## III. SYSTEM MODEL

This section presents the theoretical foundation and simulation setup for the proposed ternary signalling scheme in ICLs for 3D-ICs and details the simulation process used to evaluate its effectiveness. Different scenarios, including on-chip and off-chip crosstalk, were analysed to assess the extent of unwanted coupling. Coil structures were designed in Ansys HFSS simulator to obtain key parameters, such as Scatter Parameters (S-parameters), which were then integrated into the circuit simulator ADS. Within ADS, circuits for both NRZ and ternary signalling schemes were implemented to control current flow through the coils, allowing for a direct comparison of energy consumption, data rates, and signal integrity.

### A. Theoretical Model of ICLs

Inductive coupling in 3D-ICs enables vertical data transfer using magnetic fields generated by current pulses. According to Faraday's law of induction, a changing magnetic field within the primary coil (transmitter) induces and electomotive force (EMF) in the secondary coil (receiver), facilitating data communication across stacked layers. The relevant equations are:

$$\vec{\nabla} \times \vec{E} = -\frac{\partial \vec{B}}{\partial t} \quad \text{(Faraday's Law)} \quad (1)$$

The voltage induced in the secondary coil $V_2$ due to a changing current in the primary coil $I_1$ is given by:

$$V_2 = M \frac{dI_1}{dt} \quad (2)$$

Where $M$ represents the mutual inductance between the coils. The addition of a third voltage level (-V, 0V, +V) in ternary encoding provides a foundation for higher data throughput and enhanced signal separation, reducing the potential for interference.

### B. Simulation Setup

To validate the proposed ternary encoding scheme, the coil structures are designed within HFSS simulator, with excitations applied via a lumped port power source for computing S-parameters. S-parameters are critical for modelling the coupling between high-speed frequency signals and for analysing the coupling and voltage transfer of the coils. The HFSS simulator simulates the coils' behaviour, following the simplified model shown in Figure 2. Figure 3 shows the HFSS model with port configurations for both transmission (TX) and reception (RX) across the coils, with critical S-parameters, such as $S_{23}$ and $S_{41}$, characterizing the coupling between specific ports. The S-parameters extracted from HFSS were then imported into ADS circuit simulator where both NRZ and ternary signalling circuits were implemented. These parameters enabled ADS to simulate and compare the signal integrity, power consumption, and data transmission rates for each signaling scheme.

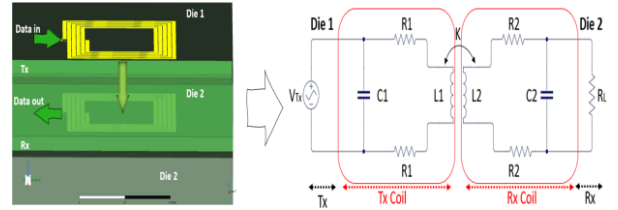

Figure 2 Simplified circuit model for an ICL

HFSS utilises voltage levels to compute s-parameters, characterising the radio-frequency components and calculating properties such as gain, loss, and other linear network parameters. These S-parameters for each coil configuration were exported and fed into the ADS circuit simulator, as illustrated in Figure 4. Table I summarizes the coil metrics used in this work, providing a comparison with values from previous studies [7][18].

Table I. Design Metrics for Inductive Coupling Coils

| Metrics | Ref [7] | Ref [18] | This work |
|---|---|---|---|
| Horizontal distance | 20um | 40 | 10 & 60um |
| Vertical distance | 300um | 88.1um | 106um |
| CMOS technology | 0.35mm | 0.35mm/65 & 28nm | 65nm |
| Inductor diameter | 100x100um | 250x250um | 250x250um |
| Adhesive | 10um | 20um | 20um |
| Number of turns | 3 | 5 | 5 |
| Array layout | 5 x 5 | 1 x 3 | 2 x 2 |
| Trace Spacing | 0.5um | 1um | 1um |
| Trace Width | 1um | 9um | 1, 3, 5um |

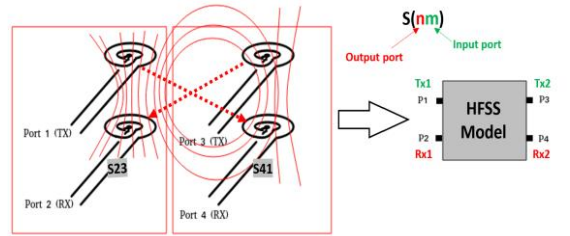

Figure 3 Explanation of S-parameters and communication scenarios for on-chip and off-chip links

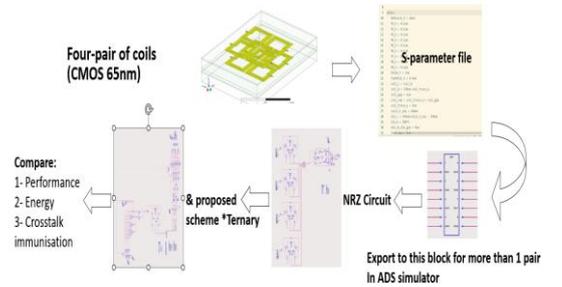

Figure 4 System Configuration Showing the Simulation Workflow

Additionally, Figure 5 presents the ternary encoding scheme used in this work, detailing the three voltage levels (-V, 0V, +V) and their corresponding binary representation. This figure illustrates how ternary encoding enables the representation of more data per symbol compared to traditional binary schemes, potentially enhancing data throughput.

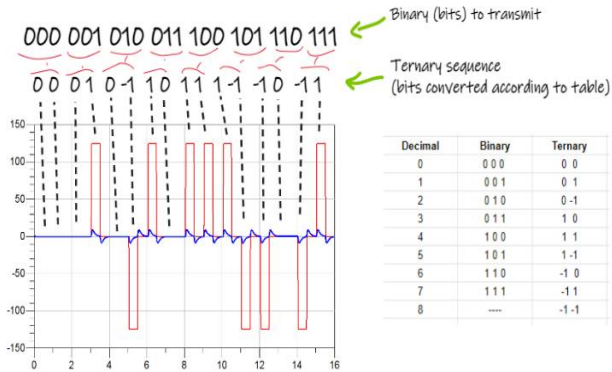

*Figure 5 Conversion of Binary Sequences to Ternary Encoding Based on The Corresponding Table, Demonstrating the Transmitted Signals*

C. Simulation Process

The simulation involves setting up current sources to simulate signal inputs to the primary coils and running the simulations to compute electromagnetic field distributions, induced voltages, and currents within the coils. These simulations are conducted using Ansys HFSS to determine the S-parameters, which are then imported into ADS simulator to be applied into NRZ and ternary circuits for further analysis. Figure 5 illustrates the overall simulation setup, showing the transmitter and receiver circuits with the HFSS model and ADS circuit simulator. The process of extracting S-parameters from HFSS and importing them into ADS for a comparative analysis of NRZ and ternary schemes is depicted in Figure 6, showing the system model configuration.

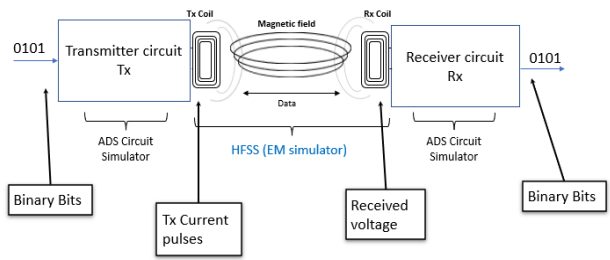

*Figure 6 Simulation Setup for Inductive Coupling Links Along with the Transmitters and Receivers Circuits*

D. Analysis and Results

Simulations were conducted to assess key performance metrics of NRZ and ternary encoding schemes, focusing on power consumption, data transmission rate, crosstalk mitigation, and symbol efficiency. The main findings are summarised below:

1- **Energy Consumption**: To evaluate energy efficiency, simulations were conducted with both 24-bit and 1,000-bit pseudo-random sequences. The NRZ encoding scheme consumed an average of 1.76 pJ for the 24-bit sequence and 97 µW for the 1,000-bit sequence. In contrast, the ternary encoding scheme demonstrated improved efficiency, with an energy requirement of only 1.2 pJ for the 24-bit sequence and 67 µW for the 1,000-bit sequence, marking a reduction of approximately 31% in power consumption across both test cases as depicted in Figure 8. This demonstrates the potential of ternary encoding to deliver substantial energy savings in high-density 3D-IC environments.

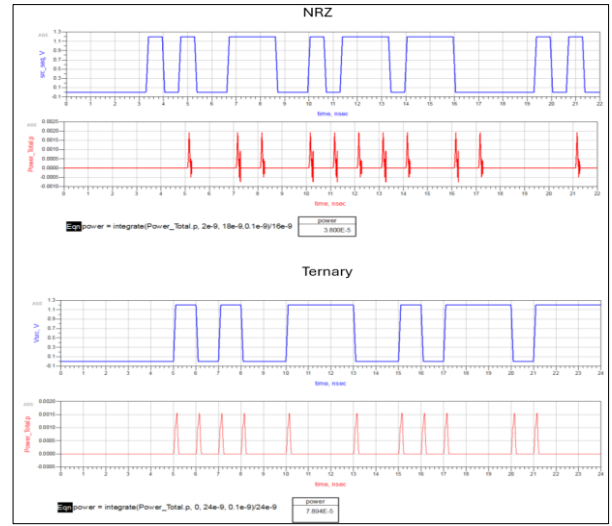

*Figure 7 Power Consumption in NRZ and Ternary Signalling*

2- **Data Transmission Rate**: The ternary scheme exhibited faster data transmission rates. As illustrated in Figure 8, the NRZ scheme took 24 ns to transmit a 24-bit sequence, whereas the ternary scheme completed the same transmission in just 16 ns, marking a 33% improvement in data rate. The proposed sequence is 000 001 010 011 100 101 110 111.

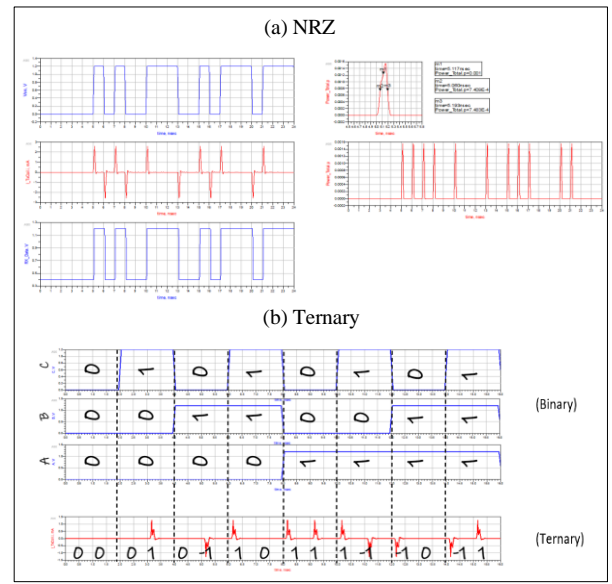

*Figure 8 NRZ vs. ternary transmission: ternary achieves 33% faster data rate with three-level encoding.*

3- **Crosstalk and Signal Integrity**: Crosstalk effects were analysed by observing current traces in the coils, particularly at the receiver side. Figure 8 shows that the reduced transition count in the ternary scheme effectively mitigated crosstalk, thereby enhancing signal integrity in stacked-layer configurations.

4- **Symbol Efficiency**: The ternary encoding scheme displayed enhanced symbol efficiency due to its three voltage levels (-1, 0, +1). This multi-level approach allowed for more data to be represented per symbol, resulting in higher data throughput with fewer transmitted pulses.

The results from the ternary scheme validate its expected advantages. By sending fewer symbols due to the 3-to-2 compression, the ternary encoding approach minimizes transitions, which enhances both data rate and power efficiency. Additionally, its design inherently supports simplified clock recovery, as a pulse is generated for each symbol combination on average. This feature, along with techniques for managing long sequences of zeros (such as adding 'stuff' bits or adjusting symbol mappings), ensures robust synchronization even under high data rates, as demonstrated by the power and transmission rate improvements over NRZ signalling.

IV. DISCUSSION

The results from this study indicate that the proposed ternary encoding scheme outperforms traditional NRZ signaling in terms of data transmission rate and power efficiency in 3D-ICs using inductive coupling links (ICLs). The observed 33% increase in data rate and 31% reduction in power consumption suggest that ternary encoding offers a viable alternative for applications requiring high data throughput and low energy consumption.

While NRZ signaling is widely used due to its simplicity, it suffers from increased power consumption and crosstalk in high-density environments. Bi-phase signaling addresses some synchronization challenges but introduces additional transitions, limiting its efficiency. Ternary encoding, by contrast, leverages three distinct voltage levels, reducing the number of transitions required per bit and achieving higher throughput. This reduction in transitions not only conserves power but also has potential benefits for signal integrity by reducing interference among neighboring coils in stacked layers.

The ternary encoding scheme's ability to achieve higher data throughput and energy efficiency is rooted in its 3-to-2 compression, which reduces transition frequency and, consequently, power consumption. The structured approach for clock recovery, ensuring a pulse per symbol combination on average, proved effective in our simulations, demonstrating reliable synchronization across scenarios. Moreover, the scheme's flexibility in managing long sequences of zeros further enhances its suitability for high-speed data transfer in 3D-IC environments, as evidenced by the observed reduction in crosstalk and improvement in signal integrity.

One consideration with ternary encoding is the added complexity in circuit design, as generating and decoding three voltage levels is more intricate than standard binary systems. The design of reliable ternary transceivers may require careful consideration to maintain signal integrity across varying environmental conditions. Additionally, integration with existing binary-based communication systems may require hybrid encoding solutions or additional circuitry, which could impact area and design complexity.

Further research could focus on optimizing ternary encoding for different 3D-IC configurations and exploring error-correction techniques to enhance robustness. Additionally, experimental validation on fabricated chips would provide more insights into practical feasibility and performance in real-world conditions.

V. CONCLUSION

In this paper, we proposed a ternary encoding scheme for inductive coupling links (ICLs) in 3D-ICs to address the limitations of traditional binary signalling methods like NRZ. The use of three voltage levels (-V, 0V, +V) allows for higher data throughput and improved power efficiency due to fewer transitions, making it well-suited for high-density 3D-IC environments. Through simulations in HFSS and ADS, we demonstrated that the ternary encoding scheme achieved a 33% increase in data transmission rate and a 31% reduction in power consumption compared to NRZ signalling. These improvements highlight the potential of ternary encoding as a robust alternative for next-generation 3D-IC applications that require efficient and reliable communication. Future work will focus on further refining the ternary encoding scheme and exploring its integration in other high-performance systems where power efficiency and signal integrity are critical.